\documentclass[a4paper,11pt]{article}
\usepackage{pos}
\usepackage{float}
\title{Probing a 96 GeV Higgs Boson in the Di-Photon Channel at the LHC}

\author[a]{R. Benbrik}
\author*[a]{M. Boukidi}
\author[b,c]{S. Moretti}
\author[b]{S. Semlali}
\affiliation[a]{Polydisciplinary Faculty, Laboratory of Fundamental and Applied Physics,\\
  Cadi Ayyad University, Sidi Bouzid, B.P. 4162, Safi, Morocco.}

\affiliation[b]{School of Physics and Astronomy,\\
University of Southampton, Southampton, SO17 1BJ, United Kingdom.}
\affiliation[c]{Department of Physics and Astronomy,\\
	Uppsala University, Box 516, SE-751 20 Uppsala, Sweden.}
\emailAdd{r.benbrik@uca.ac.ma}
\emailAdd{mohammed.boukidi@ced.uca.ma}
\emailAdd{S.Moretti@soton.ac.uk}\emailAdd{stefano.moretti@physics.uu.se}
\emailAdd{souad.semlali@soton.ac.uk}
\abstract{Recently the CMS collaboration reported a $\sim 3 \sigma$ local excess in the di-photon spectrum at 96 GeV. The same mass range concurs with a $\sim2 \sigma$ local excess in the $b\bar{b}$ invariant mass spectrum in four-jet events collected at LEP. In this contribution we show that at 1$\sigma$ level the 2HDM type-III can  perfectly fit both excesses simultaneously, while satisfying all experimental and theoretical constraints.}

\FullConference{%
  41st International Conference on High Energy physics - ICHEP2022\\
  6-13 July, 2022\\
  Bologna, Italy
}

\begin{document}
\maketitle
\section{Introduction}
The discovery of a scalar particle at the Large Hadron Collider (LHC) \cite{Aad:2012tfa,Chatrchyan:2012ufa}, has been widely hailed as the opening of a new era in particle physics. Such a Higgs boson is the first fundamental particle to be found in Nature and the last remaining undiscovered particle required for the experimental confirmation of the SM. Even though the measurements of the production and decay rates of the 125 GeV Higgs boson are currently in good agreement with the expectations within the SM, the Higgs sector chosen by Nature may not necessarily be the minimal one of such a construct. There might be additional real or complex singlets, doublets triplets, and so on, or any mixture of these. An extended Higgs sector may also be incorporated into a proper theoretical framework, for example, a specific realisation of the 2-Higgs Doublet Model (2HDM) can be made part of the Minimal Supersymmetric Standard Model (MSSM).

Light neutral Higgs bosons searches have been performed at LHC in the di-photon final state $pp\to \phi \to \gamma\gamma$. In this regard CMS has identified two excesses \cite{CMS:2018cyk} with 2$\sigma$ local significance at $m_{\gamma\gamma} = 97.6$ GeV at Run-1 (2012)  with 19.7 fb$^{-1}$ of luminosity and with 3$\sigma$ local significance at $m_{\gamma\gamma} = 95.3$ GeV at Run-2 (2016) with 35.9 fb$^{-1}$ of luminosity. Around the aforementioned mass range, an anomaly has remained from LEP data, in the $e^+e^-\to Z(H\to b\bar{b})$ channel, wherein a 2.3 $\sigma$ local excess was observed in the $b\bar{b}$ invariant mass. \cite{LEP-Excess}. 

In this work, we aim to investigate such reported excesses within a 2HDM Type-III with a specific Yukawa texture allowing for some amount of  Lepton Flavour Violation (LFV). This is done to comply with LEP, Tevatron, LHC and low energy measurements sensitive to not only the SM-like Higgs boson $H$ but also to a potentially lighter $h$ state playing the role of
the object behind the aforementioned excesses at 96 GeV. In fact,  a generic 2HDM contains five physical Higgs bosons: 2 $\mathcal{CP}$-even $h$ and $H$ ($m_h<m_H$), one $\mathcal{CP}$-odd $A$ and a pair of charged Higgs $H^\pm$. The 2HDM can be classified into Type-I, Type-II, lepton-specific, and flipped depending on how the Higgs doublets couple to the fermions. The 2HDM Type-III  corresponds to the case where each of the two Higgs doublets couples to all fermions simultaneously. Consequently, tree-level Flavour Changing Neutral Currents (FCNCs) in the sectors of charged quarks and leptons are being induced. In our approach, rather than postulating a $Z_2$ symmetry (exact or softly-broken) to control the latter, we assume a generic Yukawa texture that we will constrain by exploiting theoretical conditions  of self-consistency as well as experimental measurements of masses and couplings. 

The contribution is organized as follows: In the next section we shall introduce briefly the basic notation of 2HDM, We then move on to explain the features of the two experimental excesses (LEP and LHC). Then we will map one onto the others by presenting numerical results. We will finally conclude.
\section{General 2HDM }
The most general scalar potential of the 2HDM can be written as \cite{Branco:2011iw}:
\begin{eqnarray}
\mathcal{V} &= m_{11}^2 \Phi_1^\dagger \Phi_1+ m_{22}^2\Phi_2^\dagger\Phi_2 - \left[m_{12}^2
\Phi_1^\dagger \Phi_2 + \rm{H.c.}\right] + \lambda_1(\Phi_1^\dagger\Phi_1)^2 +
\lambda_2(\Phi_2^\dagger\Phi_2)^2 +
\lambda_3(\Phi_1^\dagger\Phi_1)(\Phi_2^\dagger\Phi_2)  ~\nonumber\\ &+
\lambda_4(\Phi_1^\dagger\Phi_2)(\Phi_2^\dagger\Phi_1) +
\frac12\left[\lambda_5(\Phi_1^\dagger\Phi_2)^2 +\rm{H.c.}\right] 
+\left\{\left[\lambda_6(\Phi_1^\dagger\Phi_1)+\lambda_7(\Phi_2^\dagger\Phi_2)\right]
(\Phi_1^\dagger\Phi_2)+\rm{H.c.}\right\} \label{C2HDMpot}
\end{eqnarray}
Adopting $\mathcal{CP}$-conserving option, and very minimal version of Higgs couplings, the above potential can be parametrized by seven free parameters, those are: Higgs masses, $m_h$, $m_H$, $m_{H^\pm}$, $m_A$, the ratio of the vacuum expectation values of the two Higgs doublets fields $\tan\beta=v_2/v_1$, the mixing angle of the $\mathcal{CP}$-even Higgs states $\alpha$, and $m^2_{12}$. In the Yukawa sector, the general scalar to fermions couplings are expressed by:
\begin{eqnarray}
-{\cal L}_Y &=& \bar Q_L Y^u_1 U_R \tilde \Phi_1 + \bar Q_L Y^{u}_2 U_R
\tilde \Phi_2  + \bar Q_L Y^d_1 D_R \Phi_1 
+ \bar Q_L Y^{d}_2 D_R \Phi_2 \nonumber \\
&+&  \bar L Y^\ell_1 \ell_R \Phi_1 + \bar L Y^{\ell}_2 \ell_R \Phi_2 + H.c. 
\label{eq:Yu}
\end{eqnarray}
where $Q_L = (u_L , d_L )$ and  $L = (\ell_L , \nu_L )$  are the doublets of $S U(2)_L$, and $Y^{f,\ell}_{1,2}$  denote the $3\times 3$ Yukawa matrices. To get naturally small FCNCs while inducing flavor violating Higgs signals, we adopt the description presented in  \cite{Cheng:1987rs} by assuming a flavour symmetry that suggest a specific texture of the Yukawa matrices, where the non-diagonal Yukawa couplings, $\tilde{Y}_{ij}$, are given in terms of fermions masses and dimensionless real parameter, $\tilde{Y}_{ij} \propto \sqrt{m_i m_j}/ v ~\chi_{ij}$.\\
After electroweak symmetry breaking (EWSB) the Yukawa Lagrangian can be expressed in terms of the mass eigenstates of the Higgs bosons, as follows:
\begin{align}
-{\cal L}^{III}_Y  &= \sum_{f=u,d,\ell} \frac{m^f_j }{v} \times\left( (\xi^f_h)_{ij}  \bar f_{Li}  f_{Rj}  h + (\xi^f_H)_{ij} \bar f_{Li}  f_{Rj} H - i (\xi^f_A)_{ij} \bar f_{Li}  f_{Rj} A \right)\nonumber\\  &+ \frac{\sqrt{2}}{v} \sum_{k=1}^3 \bar u_{i} \left[ \left( m^u_i  (\xi^{u*}_A)_{ki}  V_{kj} P_L+ V_{ik}  (\xi^d_A)_{kj}  m^d_j P_R \right) \right] d_{j}  H^+ \nonumber\\  &+ \frac{\sqrt{2}}{v}  \bar \nu_i  (\xi^\ell_A)_{ij} m^\ell_j P_R \ell_j H^+ + H.c.\, \label{eq:Yukawa_CH}
\end{align} 
Here the reduced Yukawa couplings $(\xi^{f,\ell}_\phi)_{ij}$ are given in Table~\ref{coupIII}, in terms of the mixing angles $\alpha$ and $\beta$ , and of the free parameters\footnote{The free parameters $\chi_{ij}^{f,\ell}$ are tested at the current B  physics constraints (more details can be found in Refs \cite{Benbrik:2022azi}).} $\chi_{ij}^{f,\ell}$.
\begin{table}[H]
	\begin{center}
		\setlength{\tabcolsep}{11pt}
		\renewcommand{\arraystretch}{0.6} %
		\begin{tabular}{c|c|c|c} \hline\hline 
			$\phi$  & $(\xi^u_{\phi})_{ij}$ &  $(\xi^d_{\phi})_{ij}$ &  $(\xi^\ell_{\phi})_{ij}$  \\   \hline
			$h$~ 
			& ~ $  \frac{c_\alpha}{s_\beta} \delta_{ij} -  \frac{c_{\beta-\alpha}}{\sqrt{2}s_\beta}  \sqrt{\frac{m^u_i}{m^u_j}} \chi^u_{ij}$~
			& ~ $ -\frac{s_\alpha}{c_\beta} \delta_{ij} +  \frac{c_{\beta-\alpha}}{\sqrt{2}c_\beta} \sqrt{\frac{m^d_i}{m^d_j}}\chi^d_{ij}$~
			& ~ $ -\frac{s_\alpha}{c_\beta} \delta_{ij} + \frac{c_{\beta-\alpha}}{\sqrt{2}c_\beta} \sqrt{\frac{m^\ell_i}{m^\ell_j}}  \chi^\ell_{ij}$ ~ \\
			$H$~
			& $ \frac{s_\alpha}{s_\beta} \delta_{ij} + \frac{s_{\beta-\alpha}}{\sqrt{2}s_\beta} \sqrt{\frac{m^u_i}{m^u_j}} \chi^u_{ij} $
			& $ \frac{c_\alpha}{c_\beta} \delta_{ij} - \frac{s_{\beta-\alpha}}{\sqrt{2}c_\beta} \sqrt{\frac{m^d_i}{m^d_j}}\chi^d_{ij} $ 
			& $ \frac{c_\alpha}{c_\beta} \delta_{ij} -  \frac{s_{\beta-\alpha}}{\sqrt{2}c_\beta} \sqrt{\frac{m^\ell_i}{m^\ell_j}}  \chi^\ell_{ij}$ \\
			$A$~  
			& $ \frac{1}{t_\beta} \delta_{ij}- \frac{1}{\sqrt{2}s_\beta} \sqrt{\frac{m^u_i}{m^u_j}} \chi^u_{ij} $  
			& $ t_\beta \delta_{ij} - \frac{1}{\sqrt{2}c_\beta} \sqrt{\frac{m^d_i}{m^d_j}}\chi^d_{ij}$  
			& $t_\beta \delta_{ij} -  \frac{1}{\sqrt{2}c_\beta} \sqrt{\frac{m^\ell_i}{m^\ell_j}}  \chi^\ell_{ij}$ \\ \hline \hline 
		\end{tabular}
	\end{center}
	\caption {Yukawa couplings of the $h$, $H$, and $A$ bosons to the quarks and leptons in the 2HDM Type-III.} 
	\label{coupIII}
\end{table}
\section{Explaining the excesses with 2HDM Type-III}
In this section, we investigate whether the 2HDM Type-III can explain simultaneously the excesses observed by
both LEP and CMS at 96 GeV. The evaluation of the signal strengths for these excesses was done in the narrow width approximation where we assume that the cross section ratio can be expressed via the normalized coupling:
\begin{align}
\mu_{\mathrm{LEP}}^{\mathrm{bb}}&=\frac{\sigma_{\rm 2HDM}(e^+e^-\to Zh)}{\sigma_{\rm SM}(e^+e^-\to Zh)}\cdot\frac{{\cal BR}_{\rm 2HDM}(h\to b\bar{b})}{{\cal BR}_{\rm SM}(h_{\rm SM}\to b\bar{b})}=\left|c_{hZZ}\right|^2\times \frac{{\cal BR}_{\rm 2HDM}(h\to b\bar{b})}{{\cal BR}_{\rm SM}(h_{\rm SM}\to b\bar{b})}=0.6 \pm 0.2.\label{mu_lep}\\
\mu_{\rm CMS}^{\mathrm{\gamma\gamma}}&=\frac{\sigma_{\rm 2HDM}(gg\to h)}{\sigma_{\rm SM}(gg\to h_{\rm SM})}\cdot \frac{{\cal BR}_{\rm 2HDM}(h\to \gamma\gamma)}{{\cal BR}_{\rm SM}(h_{\rm SM}\to \gamma\gamma)}=\left|c_{htt}\right|^2\times \frac{{\cal BR}_{\rm 2HDM}(h\to \gamma\gamma)}{{\cal BR}_{\rm SM}(h_{\rm SM}\to \gamma\gamma)}=0.117 \pm 0.057.\label{mu_cms}
\end{align} 	
In order to analyse whether a simultaneous fit to the observed excesses is possible, we perform a $\chi^2$ analysis where the latter  can be defined by the measured central values $\mu^{\mathrm{exp}}$ and the 1$\sigma$ uncertainties $\Delta\mu^{\mathrm{exp}}$ of the signal rates related to the two excesses as mentioned in Eq. \ref{mu_lep} and Eq. \ref{mu_cms}.
\begin{eqnarray}
\chi^2_{96}=\left(\frac{\mu_{\mathrm{LEP}}^{\mathrm{bb}}-\mu_{\mathrm{LEP}}^{\mathrm{bb, exp}}}{\Delta\mu_{\rm CMS}^{\mathrm{\gamma\gamma, exp}} }\right)^2+\left(\frac{\mu_{\rm CMS}^{\mathrm{\gamma\gamma}}-\mu_{\rm CMS}^{\mathrm{\gamma\gamma, exp}}}{\Delta\mu_{\mathrm{LEP}}^{\mathrm{bb, exp}}}\right)^2.
\end{eqnarray}

\section{Numerical Results}
Before presenting our result, we describe how theoretical self-consistency requirements and experimental measurements were used to constrain the parameter space of the 2HDM Type-III scenario that we are aiming for. The theoretical requirements  are perturbativity of the scalar quartic couplings, vacuum stability and the tree-level perturbative unitarity conditions for various scattering amplitudes of gauge and Higgs boson states (All these constraints are tested via the public code \texttt{2HDMC-1.8.0} \cite{2hdmc}). Moreover, on the experimental side, we consider constraints from electroweak precision observables (EWPO) in terms of the oblique parameters $S$, $T$, and $U$,  measurements at the LHC of the properties of the newly discovered Higgs boson (\texttt{HiggsSignal-2.6.0}\cite{HS}), Cross-section limits from  LEP, Tevatron and the LHC (\texttt{HiggsBouns-5.9.0}\cite{HB}) and finally constraints from flavor physics \cite{superIso}. For more details we refer to Ref. \cite{Benbrik:2022azi}.
Then we performed a systematic scan over the following parameter ranges.
\begin{eqnarray}
\hspace{0.7cm} m_{h} \in [80,~110]\ \text{GeV}, \ \ m_{H}= 125\ \text{GeV},
 \sin(\beta-\alpha)\in [-0.5,~-0.1], \ \  m_{A}\in [70,~90]\ \text{GeV}, \nonumber \\
 m_{H^\pm}\in [140,~180]\ \text{GeV}, \ \  \tan\beta\in [1.1,~1.5],~~m_{12}^2 = m_h^2\tan\beta/(1+\tan^2\beta).\hspace{1.15cm}
\label{parm}
\end{eqnarray}

\begin{figure}[H]	
	\centering
	\includegraphics[height=7.5cm,width=8cm]{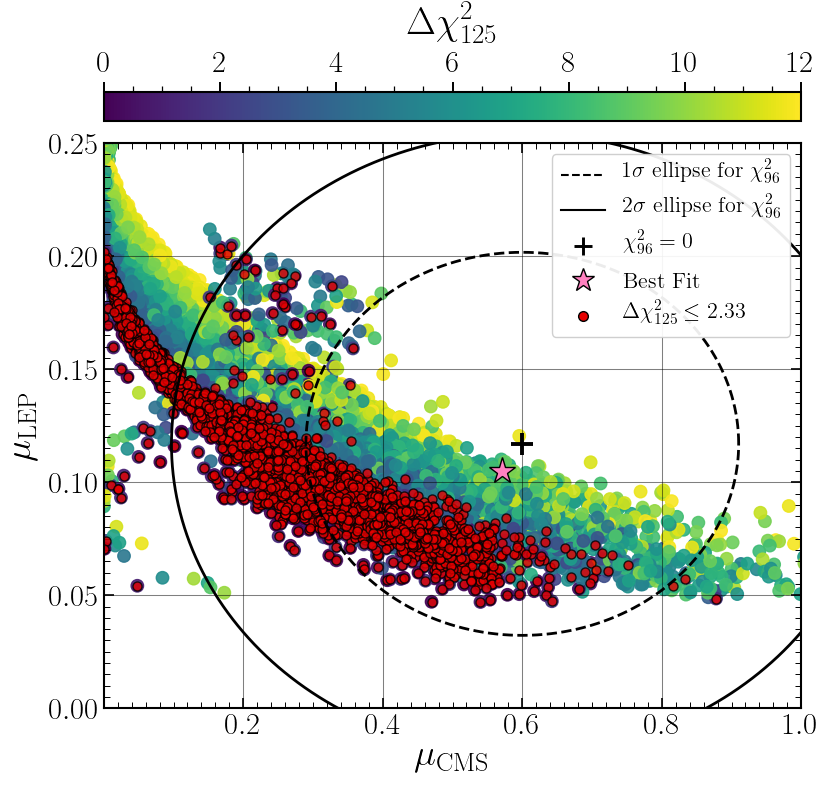}
	\caption{The signal strengths $\mu_{\rm CMS}$ and $\mu_{\mathrm{LEP}}$ following the scan described in the text. The dashed and solid black lines indicate the $1\sigma$ and $2\sigma$ ranges of $\chi^2_{96}$, respectively. The pink star corresponds to the best fit point 
		in the $h$ mass range [94, 98] GeV. The colour code indicates the $\Delta\chi^2_{125}$ values.}\label{fig3}
\end{figure}

\begin{figure}[H]
	\centering
	\includegraphics[height=7.5cm,width=14.75cm]{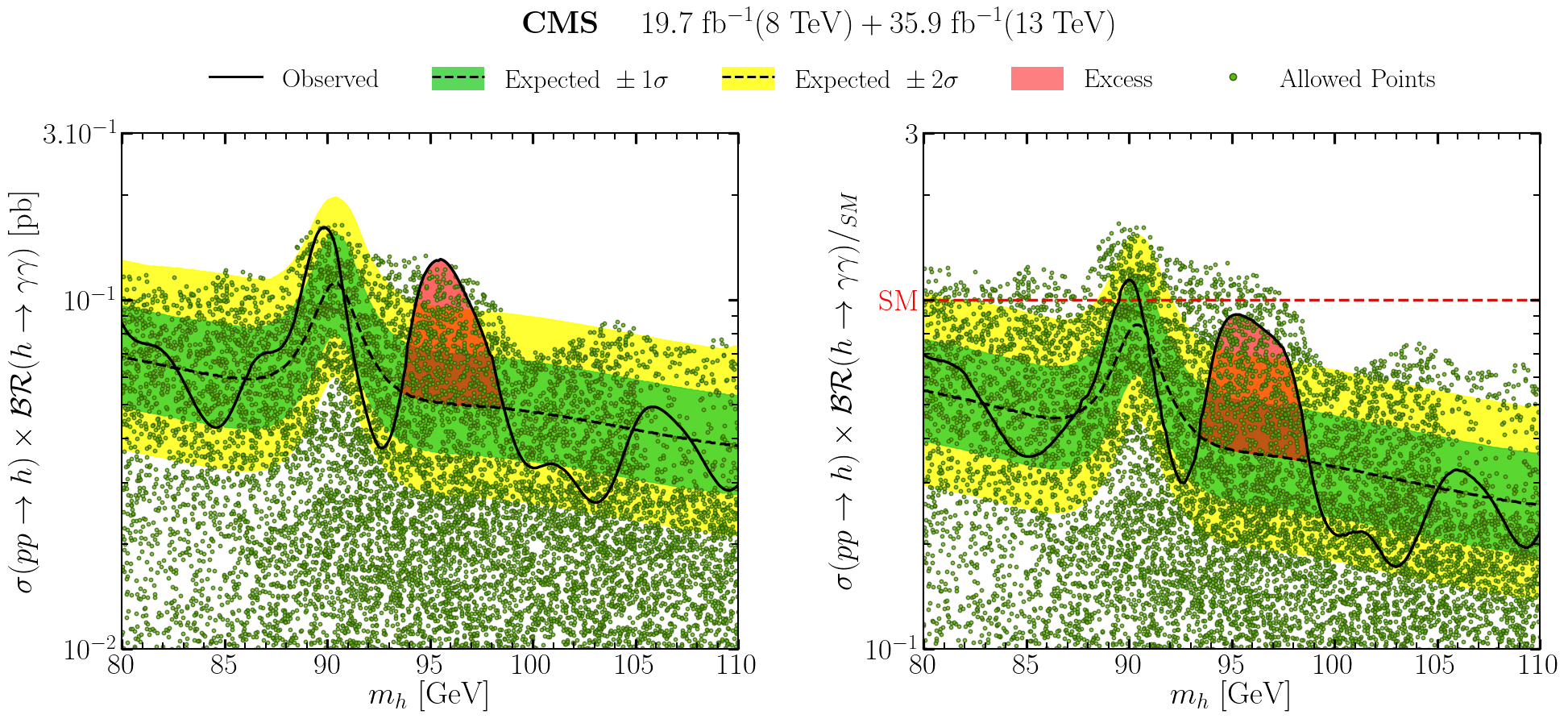}		
	\caption{Surviving points, following the discussed theoretical and experimental constraints,  
		superimposed on the results of the CMS $8+13$ TeV low-mass di-photon analysis \cite{CMS:2018cyk}. The dashed line corresponds to the expected upper limit on $\sigma \times {\cal BR}(h\rightarrow \gamma\gamma)$ (left) and $\sigma \times {\cal BR}(h\rightarrow \gamma\gamma)/\sigma^{SM} \times {\cal BR}^{SM}(h\rightarrow \gamma\gamma)$ (right) at $95\%$ C.L.,
		with $1$ and $2$ sigma errors in green and yellow, respectively. The solid line is the observed
		upper limit at $95\%$ C.L.}\label{fig4}
\end{figure}

In Fig.~\ref{fig3} we present the surviving points over the ($\mu_{\rm CMS}, \mu_{\mathrm{LEP}}$) plane, where the colour code indicates $\Delta\chi^2_{125}$. The dashed and solid lines correspond to the 1$\sigma$ and 2$\sigma$ ellipses, respectively, and the pink star indicates the best fit point. All the points shown in the figure have $\Delta\chi^2_{125} \le 12$. It is interesting to
note, that there are many points (in red) with $\Delta \chi_{125}^2\le 2.33$  within the 1$\sigma$  ellipse. Also, our best fit point is near the centre of the ellipses. Overall, the 2HDM Type-III  is fully capable of capturing the excess of 96 GeV.
Fig.~\ref{fig4} shows the predicted rate for $\sigma(pp \to h \to \gamma\gamma)$ (left) in our 2HDM Type-III and its ratio to
the SM results, $\sigma(pp \to h \to \gamma\gamma)/{\rm SM}$, for each parameter point, in combination with the expected and observed upper limits from the CMS analysis \cite{CMS:2018cyk}. Its clearly visible in both panels that many points could indeed contribute to the excess observed by CMS in the $h \to \gamma\gamma$ final state. 
\section{Conclusion}
In this contribution, we have shown how excesses observed by the CMS and LEP Collaborations could possibly be attributed to a Higgs boson produced by $gg$ fusion and decaying into $\gamma\gamma$ and $b\bar b$ with mass around 96 GeV,  we have identified the regions of parameter space where the light $\mathcal{CP}$-even state, $h$, can fit such excesses while being fully compliant with the  required signal strengths measured at both colliders. This has been accomplished after considering all the up-to-date theoretical and experimental constraints.
\section{Acknowledgments}
SM acknowledges support from the STFC Consolidated Grant ST/L000296/1 and is partially financed through the NExT Institute. The work of SS is supported in full by the NExT Institute. The work of RB and MB is supported by the Moroccan Ministry of Higher Education and Scientific Research MESRSFC and CNRST Project PPR/2015/6.

\end{document}